\documentclass[12pt]{article}
\usepackage{amssymb,amsmath,epsfig}
\allowdisplaybreaks
\begin{document}
\title{\bf Curvature-Matter Coupling Effects on Axial Gravitational Waves}
\author{M. Sharif \thanks{msharif.math@pu.edu.pk} and Aisha Siddiqa
\thanks{aisha.siddiqa17@yahoo.com}\\
Department of Mathematics, University of the Punjab,\\
Quaid-e-Azam Campus, Lahore-54590, Pakistan.}

\date{}

\maketitle
\begin{abstract}
In this paper, we investigate propagation of axial gravitational
waves in the background of flat FRW universe in $f(R,T)$ theory. The
field equations are obtained for unperturbed as well as axially
perturbed FRW metric. These field equations are solved
simultaneously to obtain the unknown perturbation parameters. We
find that the assumed perturbations can affect matter as well as
four velocity. Moreover, ignoring the material perturbations we
explicitly obtain an expression for four velocity. It is concluded
that axial gravitational waves in the curvature-matter coupling
background can produce cosmological rotation or have memory effect
if the wave profile has discontinuity at the wave front.
\end{abstract}

\section{Introduction}

The discovery of cosmic expansion is a big achievement as well as
the most fascinating area of research. Researchers introduced
different approaches to investigate the reason behind this
phenomenon by modifying matter or geometric part of the
Einstein-Hilbert action leading to modified matter models or
modified theories of gravity, respectively. Examples of modification
in geometric part are $f(R)$ \cite{1a}, $f(G)$ \cite{1b} and
$f(R,T)$ \cite{2b} theories of gravity where $R$, $G$ and $T$ denote
Ricci scalar, Gauss-Bonnet invariant and trace of the
energy-momentum tensor. While examples of modified matter models are
quintessence \cite{1d}, phantom \cite{1e}, K-essence \cite{1f},
holographic dark energy \cite{1g} and Chaplygin gas models
\cite{1h}.

The simplest generalization of general relativity (GR) is obtained
by replacing $R$ with its generic function named as $f(R)$ in the
Einstein-Hilbert action leading to $f(R)$ theory. Many astrophysical
as well as cosmological aspects have been investigated within the
framework of this theory \cite{2a}. Harko \textit{et al}. \cite{2b}
proposed $f(R,T)$ gravity which is a curvature-matter coupling
theory. This can produce a matter dependent deviation from geodesic
motion and also help to study dark energy, dark matter interactions
as well as late-time acceleration \cite{2c}.

Different aspects of cosmic and stellar evolution have been studied
in $f(R,T)$ gravity. Sharif and Zubair \cite{3a} investigated the
validity of second law of thermodynamics for phantom as well as
non-phantom phases. Shabani and Farhoudi \cite{3b} explored
viability of some $f(R,T)$ gravity models by solar system
constraints. Yousaf \textit{et al.} \cite{5h} investigated the
stability of cylindrical symmetric stellar configurations by
inducing perturbations in this theory. We have studied physical
characteristics of charged \cite{3d} as well as uncharged stellar
structure \cite{3e} in this gravity.

The fluctuations in the fabric of spacetime produced by massive
celestial objects are known as gravitational waves (GWs). The
significance of GWs comes from the fact that they lead to new
techniques to explore cosmic issues. The observations of GWs can
help us to study the individual sources of GWs that give information
about structure as well as kinematics of the cosmos. The observation
of a stochastic background of GWs of cosmological origin can provide
information about initial structure formation. These detections have
inaugurated a new era of astronomy as well as the possibility to
investigate gravity in extreme gravity regimes.

After a long history of struggles (from Weber bars to advanced laser
interferometers), scientific efforts came true and GWs are finally
detected by earth-based detectors. Some of the observed GWs signals
by the LIGO-VIRGO collaboration are GW150914 \cite{4a}, GW170104
\cite{4c} and GW170817 \cite{4d}. The origin of these signals is the
merging binaries of black holes and neutron stars which release
energy in the form of GWs. The most recent signal (GW170817)
\cite{4d} is consistent with the binary neutron star inspiral. It
has an association with gamma ray burst signal GRB170817A detected
by Fermi-GBM and provides the first direct evidence of gamma ray
bursts during the mergence of two neutron stars.

The phenomenon of GWs has become a topic of central importance in
cosmology nowadays. The polarization of a GW provides information
for its geometrical orientation. Kausar \textit{et al}. \cite{6c}
explored polarization modes of GWs in $f(R)$ theory and found two
modes other than GR. Alves \textit{et al}. \cite{6d} evaluated these
modes for $f(R,T)$ and $f(R,T^{\phi})$ theories (here $\phi$
represents scalar field). They concluded that in vacuum the former
one produces the same results as $f(R)$ while the polarization modes
in $f(R,T^{\phi})$ gravity depend upon the expression of $T^{\phi}$.
We have shown that axially symmetric dust fluid with dissipation
behave as a source of gravitational radiation in $f(R)$ theory
\cite{6a}. We have also studied polarization modes of GWs for some
viable $f(R)$ models \cite{6b}.

Regge and Wheeler \cite{5a} studied the stability of Schwarzschild
singularity by introducing small perturbations in the form of
spherical harmonics producing odd and even waves. They found that
these disturbances oscillate around equilibrium state and do not
grow with time showing the stability of Schwarzschild singularity.
Zerilli \cite{5b} analyzed the emission of gravitational radiation
when a black hole swallows a star. He did this analysis by
considering the problem of a particle falling into a Schwarzschild
black hole and perturbations introduced by Regge and Wheeler as well
as corrected the even wave propagation equation derived in
\cite{5a}. The energy carried by GWs is the gravitational radiation.
Hawking \cite{4f} investigated gravitational radiation produced by
colliding black holes and Wagoner \cite{4g} discussed these
radiation for accreting neutron stars.

Malec and Wyl\c{e}\.{z}ek \cite{5c} used the wavelike perturbations
proposed by Regge and Wheeler in the Schwarzschild spacetime to
study the GW propagation in cosmological context. They investigated
Huygens principle for cosmological GWs in Regge-Wheeler gauge and
found that this principle is satisfied in radiation dominated era
while it does not hold in matter dominated universe. Otakar
\cite{5c2} explored the GW propagations in higher dimensions using
axial perturbations proposed by Regge-Wheeler. They showed that in
braneworld scenario the Huygens principle seems to be satisfied for
high multipoles in contrast with four dimensions. Viaggiu \cite{5c3}
studied propagations of axial and polar GWs proposed in \cite{5a},
in de Sitter universe using the Laplace transformation. Kulczycki
and Malec also \cite{5d} studied the perturbations induced by axial
and polar GWs in FRW universe. They concluded that Huygens principle
has the same status for both types of waves, it is valid for
radiation era while it is broken elsewhere. The same authors
\cite{5e} discussed cosmological rotation of radiation matter
induced by axial GWs. However, axial and polar perturbations have
also been studied using gauge-invariant quantities
\cite{5e1}-\cite{5e3}. In \cite{5e3}, the authors investigated the
cosmological perturbations in the context of Lemaitre-Tolman
spacetime. In case of axial modes, their equations (restricted to
FRW metric) coincide with that of \cite{5d}.

The issues of cosmological rotation induced by GWs and validity of
Huygens principle in Regge-Wheeler gauge have not yet been studied
in the framework of modified theories. In the present work, we
induce the axial perturbations (which change the geometry from
spherical to axial) introduced by Regge and Wheeler \cite{5a} in the
flat cosmological as well as curvature-matter coupling backgrounds.
Since the FRW universes are conformally flat, these distortions are
linked with the axial GWs. These disturbances are may be the
consequence of non-gravitational forces (electromagnetic forces,
nuclear forces) associated with brutal astrophysical events. The
non-symmetric explosion of a supernova could be an example for the
production of such type of waves. We focus on the axial wave
perturbations induced in flat cosmos consisting of perfect fluid.
The paper is arranged as follows. In the coming section, we discuss
the background FRW cosmology in $f(R,T)$ theory. In section
\textbf{3}, we define the perturbations in FRW metric as well as
matter variables and formulate the corresponding field equations.
The unknown perturbation parameters are found in section \textbf{4}.
Finally, we summarize and conclude the results in the last section.

\section{FRW Cosmology and $f(R,T)$ Gravity}

In order to discuss the wave propagation in FRW universe, we
consider the FRW metric in conformal coordinates
$(\eta,r,\theta,\phi)$ as
\begin{equation}\label{1}
ds^{2}=a^{2}(\eta)(-d\eta^{2}+dr^{2}+r^{2}d\theta^{2}+r^{2}\sin^{2}\theta
d\phi^{2}),
\end{equation}
where $\eta$ is the conformal time coordinate related to the
ordinary time by the relation
\begin{equation}\label{1a}
\eta=\int\frac{dt}{a},
\end{equation}
such that the conformal Hubble parameter $H$ is related with the
ordinary Hubble parameter $\mathcal{H}$ by
\begin{equation}\label{1b}
\mathcal{H}=\frac{H}{a}.
\end{equation}
We consider matter as perfect fluid defined by the energy-momentum
tensor
\begin{eqnarray}\label{3}
T_{\mu\nu}=(\rho+p)V_{\mu}V_{\nu}+pg_{\mu\nu},
\end{eqnarray}
where $V_{\mu}$, $\rho$ and $p$ stand for four velocity, density and
pressure, respectively.

The action integral for $f(R,T)$ theory is
\begin{equation}\label{4}
S=\int d^{4}x\sqrt{-g}\left[\frac{1}{16\pi}f(R,
T)+\mathcal{L}_{m}\right].
\end{equation}
where $g$ is the determinant of the metric tensor and
$\mathcal{L}_{m}$ is the matter Lagrangian density. The field
equations for this action are
\begin{equation}\label{5}
f_{R}R_{\mu\nu}-\frac{1}{2}g_{\mu\nu}f-(\nabla_{\mu}\nabla_{\nu}-g_{\mu\nu}\Box)f_{R}=8\pi
T_{\mu\nu} -f_{T}(\Theta_{\mu\nu}+T_{\mu\nu}),
\end{equation}
where $f_{R}=\frac{\partial f}{\partial R},~f_{T}=\frac{\partial
f}{\partial T}$ and
$\Theta_{\mu\nu}=-2T_{\mu\nu}+\mathcal{L}_{m}g_{\mu\nu}$. In this
paper, we consider $f(R,T)=R+2\lambda T$ \cite{2b} to investigate
the role of curvature-matter coupling on the propagation of GWs.
This model can discuss the accelerated expansion by producing a
power-law like scale factor. It also has a correspondence with
$\Lambda$CDM model by considering the cosmological constant as a
function of trace $T$ or $\Lambda(T)$ gravity by Poplawski
\cite{r10}. The choices for matter Lagrangian density
$\mathcal{L}_{m}$ are $p$ or $-\rho$. However, it is shown that
these two densities yield the same results for minimal
curvature-matter coupling if the matter under discussion is perfect
fluid \cite{Tr1}. So the assumption $\mathcal{L}_{m}=p$ and the
model $f(R,T)=R+2\lambda T$ simplify the field equations as
\begin{equation}\label{5a}
G_{\mu\nu}=(8\pi+2\lambda) T_{\mu\nu}-2\lambda pg_{\mu\nu}+\lambda
Tg_{\mu\nu}.
\end{equation}
This yields the following independent field equations for the metric
(\ref{1})
\begin{eqnarray}\label{6}
3H^{2}=(8\pi+3\lambda)\rho_{0}a^{2}-\lambda p_{0}a^{2},\\\label{7}
-2\dot{H}-H^{2}=(8\pi+3\lambda)p_{0}a^{2}-\lambda \rho_{0}a^{2},
\end{eqnarray}
here dot denote the derivative with respect to the conformal time
$\eta$.

In further discussion, we consider the GWs in radiation dominated
era so using the equation of state (EoS) $p_{0}=\frac{\rho_{0}}{3}$,
the field equations (\ref{6}) and (\ref{7}) give the following
differential equation in $H$
\begin{eqnarray}\nonumber
2\dot{H}+\frac{6\pi+\lambda}{3\pi+\lambda}H^{2}=0,
\end{eqnarray}
which yields the scale factor
\begin{eqnarray}\label{24a}
a(\eta)=c_{1}\eta^{\frac{6\pi+2\lambda}{6\pi+\lambda}},
\end{eqnarray}
where $c_{1}$ is constant of integration. The covariant derivative
of the field equations is
\begin{eqnarray}\label{24c}
\nabla^{\mu}T_{\mu\nu}=\frac{f_{T}}{8\pi-f_{T}}\left[(T_{\mu\nu}+
\Theta_{\mu\nu})\nabla^{\mu}\ln
f_{T}+\nabla^{\mu}\Theta_{\mu\nu}-\frac{1}{2}g_{\mu\nu}\nabla^{\mu}T\right].
\end{eqnarray}
Using Eqs.(\ref{1}), (\ref{3}), the model $f(R,T)=R+2\lambda T$ and
$p_{0}=\frac{\rho_{0}}{3}$, Eq.(\ref{24c}) produces the following
differential equation in $\rho$
\begin{eqnarray}\nonumber
\dot{\rho}+3\frac{8\pi+\lambda}{6\pi+\lambda}H\rho=0,
\end{eqnarray}
whose solution is
\begin{eqnarray}\label{24b}
\rho=c_{2}a^{\frac{-3(8\pi+\lambda)}{6\pi+\lambda}},
\end{eqnarray}
$c_{2}$ is again an integration constant. These values of scale
factor and density are used in the further mathematics.

\section{Axial Perturbations in FRW Spacetime}

In this section, we first briefly discuss perturbations used to
study the effects of GWs. Here, the background metric $g_{\mu\nu}$
is the FRW spacetime and $h_{\mu\nu}$ are the corresponding
perturbations in the metric tensor due to GWs such that we have
\begin{equation}\label{1b}
g^{(perturb)}_{\mu\nu}=g^{(flat)}_{\mu\nu}+eh_{\mu\nu}+O(e^{2}),
\end{equation}
where $e$ is a small parameter (it measures strength of
perturbations and the terms involving $O(e^{2})$ are neglected).

We follow the Regge-Wheeler \cite{5a} perturbation scheme to
investigate the wavelike fluctuations. To obtain explicit
expressions for the components of $h_{\mu\nu}$ in terms of four
coordinates ($x^{0}=\eta,~x^{1}=r,~x^{2}=\theta,~x^{3}=\phi$), they
expressed them in the form of spherical harmonics. The symmetry of
the metric tensor allows the angular momentum to be defined. The
angular momentum is discussed by assuming the rotations on a
2D-manifold with $\eta=$ constant and $r=$ constant. The components
of $h_{\mu\nu}$ have different transformations under a rotation of
the frame. Among the ten independent components of the tensor
$h_{\mu\nu}$, the components $h_{00},~h_{01},~h_{11}$ transform like
scalars (as $x^{0}=\eta$ and $x^{1}=r$ are constants and do not
change during rotation), $h_{02},~h_{03},~h_{12},~h_{13}$ change
like vectors (as $x^{2}$ and $x^{3}$ are changed during rotation)
while $h_{22},~h_{23},~h_{33}$ transform like tensors. Further,
these scalars, vectors and tensors are expressed in terms of
spherical harmonics $Y_{L}^{~M}$ where $L$ is the angular momentum
with the projection $M$ on $z$-axis. After this, they expressed the
perturbation matrix $h_{\mu\nu}$ in terms of odd and even parity
waves. In this paper, we only consider the odd or axial wave
perturbations defined by the matrix \cite{5a}
\begin{equation}
h_{\mu\nu}=\partial_{\theta}Y\sin\theta \left(
\begin{array}{cccc}
0 & 0 & 0 & k_{0} \\
0 & 0 & 0 & k_{1} \\
0 & 0 & 0 & 0 \\
k_{0} & k_{1} & 0 & 0 \\
\end{array}
\right),
\end{equation}
with $k_{0}=k_{0}(\eta,r)$ and $k_{1}=k_{1}(\eta,r)$. Here we are
considering the odd waves corresponding to $m=0$, which are
discussed by Regge and Wheeler \cite{5a} so that $\phi$ disappears
in calculations. Also, for the wavelike solution the index $l$
exceeds one, i.e., $Y=Y_{l0};~l=2,3,...$. The resulting axially
perturbed FRW spacetime in Regge-Wheeler gauge is defined by
\begin{eqnarray}\nonumber
ds^{2}&=&-a^{2}(\eta)d\eta^{2}+2ek_{0}\partial_{\theta}Y\sin\theta
d\eta d\phi+a^{2}(\eta)dr^{2}+2ek_{1}\partial_{\theta}Y\sin\theta
drd\phi\\\label{2}&+&a^{2}(\eta)r^{2}d\theta^{2}+a^{2}(\eta)r^{2}\sin^{2}\theta
d\phi^{2}+O(e^{2}).
\end{eqnarray}

The perturbations in the material quantities are defined as follows
\cite{5d}
\begin{eqnarray}\label{3b}
\rho&=&\rho_{0}(1+e\Delta(\eta,r)Y)+O(e^{2}),\\\label{3b1}
p&=&p_{0}(1+e\Pi(\eta,r)Y)+O(e^{2}),
\end{eqnarray}
where $\rho_{0}$ and $p_{0}$ are the background density and
pressure. The fluid may or may not be comoving in the perturbed
scenario so the perturbed components of four velocity are taken as
\cite{5d}
\begin{eqnarray}\label{v1}
V_{0}&=&\frac{2g^{(0)}_{00}+ek_{00}}{2a(\eta)}+O(e^{2}),\\\label{v2}
V_{1}&=&ea(\eta)w(\eta,r)Y+O(e^{2}),\\\label{v3}
V_{2}&=&ev(\eta,r)Y'+O(e^{2}),\\\label{v4} V_{3}&=&e\sin\theta
u(\eta,r)Y'+O(e^{2}),
\end{eqnarray}
where $V_{\alpha}V^{\alpha}=-1+O(e^{2})$. The field equations for
the perturbed metric (\ref{2}) as well as corresponding perturbed
matter are
\begin{eqnarray}\label{8}
&&3H^{2}=[(8\pi+3\lambda)\rho_{0}-\lambda
p_{0}+(8\pi+3\lambda)\rho_{0}e\Delta Y-\lambda p_{0}e\Pi
Y]a^{2},\\\label{9}
&&w(8\pi+2\lambda)(\rho_{0}+p_{0})=0,\\\label{10}
&&v(8\pi+2\lambda)(\rho_{0}+p_{0})=0,\\\nonumber
&&-2\dot{H}-H^{2}=a^{2}[(8\pi+3\lambda)p_{0}-\lambda
\rho_{0}+(8\pi+3\lambda)p_{0}e\Pi Y-\lambda \rho_{0}e\Delta
Y],\\\label{11} \\\label{12} &&k'_{1}=\dot{k}_{0},\\\nonumber
&&\dot{k}'_{1}-k''_{0}+\frac{2}{r}\dot{k}_{1}-2Hk'_{1}+\frac{4}{r}k_{1}H-4k_{0}\dot{H}-2H^{2}k_{0}+
\frac{k_{0}}{r^{2}}l(l+1)\\\nonumber&&=-2a^{3}(8\pi+2\lambda)u(\rho_{0}+p_{0})+
[(8\pi+3\lambda)p_{0}-\lambda\rho_{0}]2a^{2}k_{0}\\\label{13}&&+2a^{2}e(8\pi+4\lambda)k_{0}p_{0}\Pi
Y-2a^{2}e\lambda k_{0}\rho_{0}\Delta,\\\nonumber
&&\ddot{k}_{1}-\dot{k}'_{0}+\frac{2}{r}\dot{k}_{0}-\frac{2}{r^{2}}k_{1}
-2H\dot{k}_{1}-6\dot{H}k_{1}-2H^{2}k_{1}+\frac{k_{1}}{r^{2}}l(l+1)
\\\nonumber&&=2a^{2}k_{1}p_{0}(8\pi+2\lambda)+2\lambda
a^{2}k_{1}(-\rho_{0}+p_{0})-2a^{2}e\lambda
k_{1}\rho_{0}\Delta\\\label{14}&&+2a^{2}e(8\pi+4\lambda)k_{1}p_{0}\Pi
Y,
\end{eqnarray}
where prime indicates the derivative with respect to $r$ and also,
we have used the relation \cite{5d}
\begin{equation}\nonumber
\partial_{\theta}\partial_{\theta}Y=-l(l+1)Y-\cot\theta
\partial_{\theta}Y.
\end{equation}

\section{Effects of Axial Gravitational Waves}

In this section, we find expressions for the perturbation parameters
$k_{0}$, $k_{1}$, $\Delta$, $\Pi$, $w$, $v$ and $u$. Equation
(\ref{9}) implies that either the factor $(8\pi+2\lambda)=0$, i.e.,
$\lambda=-4\pi$ or $w(\rho_{0}+p_{0})=0$. However, the viability
conditions for $f(R,T)$ gravity models are
\begin{equation}\nonumber
f_{R}>0,\quad 1+\frac{f_{T}}{8\pi}>0 \quad \text{and}\quad f_{RR}>0,
\end{equation}
and give the constraint $\lambda>-4\pi$ for our model implying that
$(8\pi+2\lambda)\neq0$. Hence Eqs.(\ref{9}) and (\ref{10}) yield
that $w=0$ and $v=0$. Substituting the unperturbed field equations
in perturbed one, we obtain the following equations from (\ref{8}),
(\ref{11}), (\ref{13}) and (\ref{14}), respectively.
\begin{eqnarray}\label{15}
&&(8\pi+3\lambda)\rho_{0}\Delta-\lambda p_{0}\Pi =0,\\\label{16}
&&(8\pi+3\lambda)p_{0}\Pi-\lambda\rho_{0}\Delta=0,\\\nonumber
&&\dot{k}'_{1}-k''_{0}+\frac{2}{r}\dot{k}_{1}-2Hk'_{1}+\frac{4}{r}k_{1}H+
\frac{k_{0}}{r^{2}}l(l+1)\\\label{17}&&=-2a^{3}(8\pi+2\lambda)u(\rho_{0}+p_{0})
+2a^{2}e[(8\pi+4\lambda)k_{0}p_{0}\Pi Y-\lambda
k_{0}\rho_{0}\Delta],\\\nonumber
&&\ddot{k}_{1}-\dot{k}'_{0}+\frac{2}{r}\dot{k}_{0}-\frac{2}{r^{2}}k_{1}
-2H\dot{k}_{1}-2\dot{H}k_{1}+\frac{k_{1}}{r^{2}}l(l+1)\\\label{18}&&=2a^{2}e[-\lambda
k_{1}\rho_{0}\Delta+(8\pi+4\lambda)k_{1}p_{0}\Pi Y].
\end{eqnarray}
Solving (\ref{15}) and (\ref{16}) simultaneously for $\Pi$, we
obtain
\begin{eqnarray}\label{19}
((8\pi+3\lambda)^{2}-\lambda^{2}) p_{0}\Pi =0,
\end{eqnarray}
which implies either
\begin{eqnarray}\label{22}
((8\pi+3\lambda)^{2}-\lambda^{2})=0\quad \text{or} \quad \Pi =0.
\end{eqnarray}

The first factor in the above equation yields $\lambda=-4\pi$ and
$-2\pi$. However, keeping in mind the viability conditions for the
assumed model, we exclude $\lambda=-4\pi$. Hence if $\lambda=-2\pi$,
then there is a possibility that $\Pi\neq0$ and similarly
$\Delta\neq0$, i.e., the axial GWs can affect the background matter
in curvature-matter coupling scenario. Assuming the EoS for
radiation dominated era $p_{0}=\frac{1}{3}\rho_{0}$, we obtain the
following relationship between $\Pi$ and $\Delta$
\begin{eqnarray}\label{22a}
\Pi=3\left(\frac{8\pi}{\lambda}+3\right)\Delta.
\end{eqnarray}
Substituting the above relation in Eqs.(\ref{17}) and (\ref{18}), we
are left with four unknowns $k_{0}$, $k_{1}$, $\Delta$, $u$ with
three equations (\ref{12}), (\ref{17}), (\ref{18}). Thus in order to
have the system closed, we assume that GWs do not perturb the matter
field, i.e., $\Delta=0=\Pi$. Now introducing a new quantity
$Q(\eta,r)$ such that
\begin{eqnarray}\label{23}
k_{1}(\eta,r)=ra(\eta)Q(\eta,r).
\end{eqnarray}
Using this equation with Eq.(\ref{12}) in (\ref{18}), we obtain
\begin{eqnarray}\label{24}
\ddot{Q}-Q''+\frac{l(l+1)}{r^{2}}Q-a^{2}[\frac{(4\pi+3\lambda)}{3}\rho_{0}
-\frac{(12\pi+5\lambda)}{3}p_{0}]Q=0.
\end{eqnarray}

Inserting $p_{0}=\frac{\rho_{0}}{3}$, the values of $a(\eta)$ as
well as $\rho_{0}$ from Eqs.(\ref{24a}) and (\ref{24c}) into
(\ref{24}), it follows that
\begin{eqnarray}\label{25}
\ddot{Q}-Q''+[\frac{l(l+1)}{r^{2}}-\frac{4c_{1}^{2}c_{2}b_{1}\lambda}{9}
\eta^{\frac{6\pi+2\lambda}{6\pi+\lambda}\frac{(-3)(8\pi+\lambda)}{6\pi+\lambda}}]Q=0,
\end{eqnarray}
where $b_{1}=c_{1}^{\frac{-3(8\pi+\lambda)}{6\pi+\lambda}}$. Let us
define $A=\frac{4c_{1}^{2}c_{2}b_{1}\lambda}{9}$ and take $l=2$ such
that the above equation becomes
\begin{eqnarray}\label{26}
\ddot{Q}-Q''+[\frac{6}{r^{2}}-A\eta^{\frac{6\pi+2\lambda}
{6\pi+\lambda}\frac{(-3)(8\pi+\lambda)}{6\pi+\lambda}}]Q=0.
\end{eqnarray}
This is a wave equation and can be solved through separation of
variables by assuming $Q(\eta,r)=\mathcal{T}(\eta)\mathcal{R}(r)$
and the initial conditions.
\begin{equation}\nonumber
Q(0,r)=\Psi_{1}(r),\quad
\partial_{\eta}Q(0,r)=\Psi_{2}(r),
\end{equation}

Introducing the separation constant $-m^{2}$, we obtain the
following two differential equations
\begin{eqnarray}\label{27}
\ddot{\mathcal{T}}-\left(A\eta^{\frac{6\pi+2\lambda}
{6\pi+\lambda}\frac{(-3)(8\pi+\lambda)}{6\pi+\lambda}}-m^{2}\right)\mathcal{T}&=&0,\\\label{28}
\mathcal{R}''-\left(\frac{6}{r^{2}}-m^{2}\right)\mathcal{R}&=&0.
\end{eqnarray}

These are second order homogeneous linear differential equations
with variable coefficients. Equation (\ref{27}) can yield some
solution if the power of $\eta$ is fixed. So, we consider
$\frac{6\pi+2\lambda}
{6\pi+\lambda}\frac{(-3)(8\pi+\lambda)}{6\pi+\lambda}=n$ and check
that for what values of $n$, the values of $\lambda$ are consistent
with viability criteria. We find that the values of $\lambda$ for
$n>1$ are not consistent with $\lambda>-4\pi$ (the viability
criteria) and $n<-2$ yields imaginary values of $\lambda$. Hence,
$n$ can have the values within the limit $-2\leq n<1$. For $n=-2$,
we have $\lambda=0$ which is the case of GR. For convenience, we
consider the integer values in this interval, i.e., $n=0,-1$, to
find the solution of Eq.(\ref{27}). For $n=0$, the solution is
\begin{equation}
\mathcal{T}(\eta)=c_{3}\cos m\eta+c_{4} \sin m\eta,
\end{equation}
where $c_{3}$ and $c_{4}$ are constants of integration and for
$n=-1$, we have
\begin{eqnarray}\nonumber
\mathcal{T}(\eta)&=&c_{5}\eta
e^{-im\eta}\text{Hypergeometric1F1}\left[1+\frac{A}{2im},2,2im\eta\right]\\\nonumber&+&c_{6}
\eta
e^{-im\eta}\text{HypergeometricU}\left[1+\frac{A}{2im},2,2im\eta\right],
\end{eqnarray}
where $c_{5}$, $c_{6}$ are constants and Hypergeometric1F1,
HypergeometricU are the confluent hypergeometric functions of the
first and second kind, respectively. These functions are defined by
\begin{eqnarray}\nonumber
\text{Hypergeometric1F1}(\alpha;\beta;z)&=&\frac{\Gamma(\beta)}{\Gamma(\beta-\alpha)\Gamma(\alpha)}
\int_{0}^{1}e^{zt}t(\alpha-1)(1-t)^{\beta-\alpha-1}dt,\\\nonumber
\text{HypergeometricU}(\alpha,\beta,z)&=&\frac{1}{\Gamma(\alpha)}
\int_{0}^{\infty}e^{-zt}t(\alpha-1)(1+t)^{\beta-\alpha-1}dt,
\end{eqnarray}
where $``\Gamma"$ indicates the gamma function. The solution of
Eq.(\ref{28}) is obtained as
\begin{eqnarray}\nonumber
\mathcal{R}(r)&=&\sqrt{\frac{2}{m\pi}}c_{7}\left(\frac{-3\cos
mr}{mr}-\sin mr+\frac{3\sin
mr}{m^{2}r^{2}}\right)\\\label{27a}&+&\sqrt{\frac{2}{m\pi}}c_{8}\left(\frac{-3\cos
mr}{m^{2}r^{2}}-\frac{3\sin mr}{mr}+\cos mr\right),
\end{eqnarray}
where $c_{7}$, $c_{8}$ are integration constants. Inserting the
values of $\mathcal{R}(r)$ and $\mathcal{T}(\eta)$ in
$Q(\eta,r)=\mathcal{T}(\eta)\mathcal{R}(r)$, we obtain $Q(\eta,r)$
for both values of $n$. Furthermore, using initial conditions one
can find the expressions for $\Psi_{1}(r)$ and $\Psi_{2}(r)$ for
$n=0$ as well as $n=-1$.

Replacing the values of $Q(\eta,r)$ and $a(\eta)$ in Eq.(\ref{23}),
we obtain the value of $k_{1}$ while the expression for $k_{0}$ is
obtained from Eq.(\ref{12}) as follows
\begin{eqnarray}\label{29}
k_{0}=B(r)+\int_{\eta_{0}}^{\eta} k'_{1}(\tau,r)d\tau,
\end{eqnarray}
where $\eta_{0}$ is the conformal time at the hypersurface
originating GWs. Assuming $k_{0}(\eta,r)=0$, we have $B(r)=0$ and
$k_{0}$ becomes
\begin{eqnarray}\label{29a}
k_{0}=(r\mathcal{R}(r))'\int_{\eta_{0}}^{\eta}
a(\tau)\mathcal{T}(\tau)d\tau.
\end{eqnarray}
Finally, replacing the values of $k_{0}$, $k_{1}$ and $\Delta=0=\Pi$
in Eq.(\ref{17}), we obtain for $n=0$
\begin{eqnarray}\nonumber
&&u(\eta,r)=\frac{\eta^{\frac{\lambda}{6\pi+\lambda}}c_{1}}{(mr)^{\frac{5}{2}}}
\sqrt{\frac{2r}{\pi}}[(3mrc_{7}+m^{3}r^{3}c_{7}+3c_{8})\cos
mr+(-3c_{7}+mr\\\label{30a}&&(3+m^{2}r^{2})c_{8})\sin
mr][(c_{3}-m\eta c_{4})\cos m\eta+(m\eta c_{3}+c_{4})\sin m\eta].
\end{eqnarray}
For $n=-1$, we have
\begin{eqnarray}\nonumber
u(\eta,r)&=&\frac{\eta^{\frac{6\pi+2\lambda}{6\pi+\lambda}}e^{-im\eta}c_{1}}{(mr)^{\frac{5}{2}}}
\sqrt{\frac{r}{2\pi}}\left[(1-\frac{6\pi+2\lambda}{6\pi+\lambda}+im\eta)
\left\{2c_{5}\right.\right.\\\nonumber&\times&\left.\left.\text{Hypergeometric1F1}\left[1+\frac{A}{2im},2,2im\eta\right]
+2c_{6}\right.\right.\\\nonumber&\times&\left.\left.\text{HypergeometricU}\left[1+\frac{A}{2im},2,2im\eta\right]\right\}
+(A+2im)\eta\right.\\\nonumber&\times&\left.
\left\{2c_{6}\text{HypergeometricU}\left[2+\frac{A}{2im},3,2im\eta\right]
-c_{5}\right.\right.\\\nonumber&\times&\left.\left.\text{Hypergeometric1F1}
\left[2+\frac{A}{2im},3,2im\eta\right]\right\}\right][(3mrc_{7}+m^{3}r^{3}c_{7}\\\label{30b}&+&3c_{8})\cos
mr+(-3c_{7}+mr(3+m^{2}r^{2})c_{8})\sin mr].
\end{eqnarray}
Thus the final expression for four velocity in radiation dominated
phase becomes
\begin{eqnarray}
V_{\alpha}=(-c_{1}\eta^{\frac{6\pi+2\lambda}{6\pi+\lambda}},0,0,
e\partial_{\theta}Yu(\eta,r)\sin\theta).
\end{eqnarray}
Also,
$Y=Y_{20}(\theta)=\frac{1}{4}\sqrt{\frac{5}{\pi}}(3\cos^{2}\theta-1)$
leads to
$\partial_{\theta}Y=\frac{1}{4}\sqrt{\frac{5}{\pi}}\cos\theta\sin\theta$
and hence
\begin{eqnarray}
V_{\alpha}=(-c_{1}\eta^{\frac{6\pi+2\lambda}{6\pi+\lambda}},0,0,
\frac{e}{4}\sqrt{\frac{5}{\pi}}u(\eta,r)\cos\theta\sin^{2}\theta).
\end{eqnarray}
Thus the azimuthal velocity of any point P having coordinates
$(\eta,r,\theta,\phi)$ is
$V_{3}=\frac{e}{4}\sqrt{\frac{5}{\pi}}u(\eta,r)\cos\theta\sin^{2}\theta$,
where $u(\eta,r)$ is given in Eq.(\ref{30a}) and (\ref{30b}) for
$n=0,-1$, respectively.

\section{Final Remarks}

According to rough approximate, a pair of massive black holes merge
in every $223_{-115}^{352} sec$ and a binary of neutron star merge
in every $13_{-9}^{49} sec$ \cite{c1}. Among these mergers a small
fraction is detected by advance interferometers of LIGO-Virgo
collaboration and can be associated to some individual GW event. The
rest of the events contribute to make a stochastic background which
is a random GW signal originated by various independent, weak and
unresolved sources. These sources include for instance, the
supernova explosions at the end of a massive star's life (including
non-symmetric explosions), a rapidly rotating neutron star, cosmic
strings etc. Mathematical and statistical approaches have been
developed to observe these stochastic background of GWs and extract
information from them \cite{c2}. These GWs signals have great
influence on cosmic evolution and hence the study of different
aspect of GW phenomenon is very significant.

The main goal of this manuscript is to explore the changes produced
by axial GWs in geometry as well as matter of a flat universe during
evolution and in the context of curvature-matter coupling theory.
For this purpose, we assume the presence of these waves and find the
corresponding geometrical and material changes produced by these
waves in $f(R,T)$ gravity. We have introduced axial perturbations in
the flat FRW spacetime, the background matter is also perturbed as
well as the four velocity is allowed to be non-comoving. We then
proceed to find all unknown parameters of perturbations with the
help of perturbed and unperturbed field equations. It is mentioned
here that all field equations reduce to GR equations \cite{5e} for
$\lambda=0$.

The factors $w$, $v$, appearing in $V_{1}$ and $V_{2}$ are zero
showing that axial waves do not change these components of velocity
which is similar to that in GR. We have found that axial GWs in
$f(R,T)$ theory can perturb the background matter in contrast to GR.
However, here we suppose $\Delta$ and $\Pi$ equal to zero in order
to find the remaining functions $k_{0}$, $k_{1}$ and $u$. The
resulting $k_{0}$ and $k_{1}$ are different from those of GR and
depend upon the coupling constant $\lambda$. The function $u$
appearing in the azimuthal velocity component has non-zero
expression showing that fluid exhibits a rotation due to axial GWs
similar to GR. But the expression of $u$ here depends upon $\lambda$
and differs from GR.

Currently, our universe is in expansion phase and it is crucial to
investigate the propagation of GWs in this expanding universe. In
this regard, we expand our analysis using the EoS $p_{0}=-\rho_{0}$
for expanding matter and observe how such types of GWs can perturb
the flat cosmos in the recent era. For $p_{0}=-\rho_{0}$, the scale
factor and density have the expressions
\begin{equation}
a(\eta)=\frac{\tilde{c}_{1}}{\eta},\quad
\rho=\tilde{c}_{2}a^{\frac{-3\lambda}{2\pi}},
\end{equation}
where $\tilde{c}_{1}$ and $\tilde{c}_{2}$ are integration constants.
It is found that this EoS can yield non-vanishing $w$ and $v$ (from
(\ref{9}) and (\ref{10})) while the remaining expressions remain the
same with $A=\frac{8(2\pi+\lambda)}{3}c_{1}^{2}c_{2}b_{1}$,
$b_{1}=\tilde{c}_{6}^{\frac{-3\lambda}{2\pi}}$ and
$\frac{3\lambda-4\pi}{2\pi}=n$. In dark energy dominated phase, $n$
can take positive and negative values, however, similar to radiation
dominated phase, $n=-2$ yields the GR case.

The angular ($\Omega$) and linear rotational ($V$) velocities of the
fluid are
\begin{eqnarray}\nonumber
\Omega&=&\frac{V^{3}}{V^{0}}=\frac{Cu}{ar^{2}}\cos\theta; \quad
C=\frac{e}{4}\sqrt{\frac{5}{\pi}},\\\nonumber V&=&ar\sin\theta
\Omega=\frac{Cu\sin 2\theta}{2r}.
\end{eqnarray}
When the expression of $u(\eta,r)$ is continuous at the wave front,
the smooth wave profile does not induce any cosmological rotation
\cite{5e}. Hence we conclude that the axial GW can induce a
cosmological rotation if $u(\eta,r)$ is discontinuous at the wave
front. If the freely falling particles are displaced by a GW, it is
called memory effect of the GW. Hence the axial GW in $f(R,T)$
gravity induces memory effect when the wave profile has
discontinuity at the wave front. Also, the model considered here
describes the simplest curvature-matter coupling and we assume this
model to reduce the calculation work. However, this work can be
extended for other minimally coupled models containing nonlinear
power of $R$ or $T$ or non-minimally coupled models leading to
interesting results. Such models may yield the non-vanishing values
of the perturbation parameters which are zero in the present
scenario.

When a GW without memory passes through a detector, it produces an
oscillatory deformation and returns the detector back to its
equilibrium state. On the other hand, a GW with memory can induce a
permanent deformation in an idealized detector, i.e., a truly free
falling detector \cite{Tr3}. The detectors like Weber bars and LIGO
are not sensitive to the memory effect. However, the detectors of
the type like LISA (Laser interferometry space antenna) or advanced
LIGO can detect the memory due to its sensitivity and with strong
memory sources \cite{Tr4}. Also, the ground-based detectors are not
truly free falling and cannot store a memory signal while LISA like
detectors are able to maintain the permanent displacement because
these are free floating.

\vspace{1.0cm}

{\bf Acknowledgment}

\vspace{0.25cm}

We would like to thank the Higher Education Commission, Islamabad,
Pakistan for its financial support through the {\it Indigenous Ph.D.
5000 Fellowship Program Phase-II, Batch-III}.

\end{document}